\documentclass[journal,10pt]{IEEEtran}
\IEEEoverridecommandlockouts  \usepackage{cite}
\usepackage{amsmath,amssymb,amsfonts}
\usepackage{setspace}
\usepackage{graphicx}
\usepackage{paralist}
\usepackage{textcomp}
\usepackage{mathtools}
\usepackage{xcolor}
\usepackage{color}
\usepackage{algorithm}
\usepackage{algpseudocode}
\usepackage{mathabx} \usepackage{optidef}
\usepackage{tcolorbox}
\usepackage{multirow}
\usepackage{lipsum}
\usepackage{verbatim}
\usepackage{afterpage}
\usepackage[normalem]{ulem}
\usepackage{dsfont}
\usepackage{listings}
\usepackage{url}
\usepackage[utf8]{inputenc}
\graphicspath{ {./images/} }
\usepackage{acronym}
\usepackage{cite}
\usepackage{lettrine}
\usepackage{subfigure}
\usepackage{enumitem}
\usepackage{array}
\usepackage{amsthm}

\def\BibTeX{{\rm B\kern-.05em{\sc i\kern-.025em b}\kern-.08em
    T\kern-.1667em\lower.7ex\hbox{E}\kern-.125emX}}
    \newcolumntype{M}[1]{>{\centering\arraybackslash}m{#1}}
\begin{document}
\setstretch{1}
\title{\fontsize{22.8}{27.6}\selectfont UAV-Based Cell-Free Massive MIMO: 

Joint Placement and Power Optimization under Fronthaul Capacity Limitations}
\author{Neetu R.R, Ozan Alp Topal, Özlem Tuğfe Demir, Emil Björnson, Cicek Cavdar, Gourab Ghatak and Vivek Ashok Bohara
\thanks{Neetu R.R. and Vivek Ashok Bohara are affiliated with the Department of Electronics and Communication Engineering, IIIT-Delhi, India. (Email: {neetur, vivek.b}@iiitd.ac.in). Ozan Alp Topal, Emil Björnson, and Cicek Cavdar are affiliated with the Department of Computer Science, KTH Royal Institute of Technology, Kista, Sweden. (Email: {oatopal, emilbjo, cavdar}@kth.se). Özlem Tuğfe Demir is with the Department of Electrical and Electronics
Engineering, TOBB University of Economics and Technology, Ankara,
Türkiye (Email: ozlemtugfedemir@etu.edu.tr). Gourab Ghatak is with the Department of Electrical Engineering, IIT Delhi, India. (Email: gghatak@ee.iitd.ac.in). 
This research is funded by the IIT Palakkad Technology IHub Foundation Doctoral Fellowship IPTIF/HRD/DF/026.
}}

\acrodef{UAV}[UAV]{unmanned aerial vehicle}
\acrodef{MU-MIMO}[MU-MIMO]{multi user-MIMO}
\acrodef{UAV-AP}[UAV-AP]{UAV-access point}
\acrodef{UE}[UE]{user equipment}
\acrodef{6G}[6G]{sixth-generation}
\acrodef{C-RAN}[C-RAN]{centralized-RAN}
\acrodef{RAN}[RAN]{random access network}
\acrodef{CPU}[CPU]{central processing unit}
\acrodef{BBU}[BBU]{baseband unit}
\acrodef{RRH}[RRH]{remote radio head}
\acrodef{GOPS}[GOPS]{giga-operations per second}
\acrodef{3GPP}[3GPP]{$3^{rd}$ Generation Partnership Project}
\acrodef{CU}[CU]{centralized unit}
\acrodef{DU}[DU]{distributed unit}
\acrodef{RU}[RU]{radio unit}
\acrodef{RF}[RF]{radio frequency}
\acrodef{QoS}[QoS]{quality of service}
\acrodef{CSI}[CSI]{channel state information}
\acrodef{SINR}[SINR]{signal-to-interference-plus-noise ratio}
\acrodef{FDMA}[FDMA]{frequency division multiple access}
\acrodef{SDMA}[SDMA]{spatial division multiple access}
\acrodef{LoS}[LoS]{line-of-sight}
\acrodef{NLoS}[NLoS]{non line-of-sight}
\acrodef{mmWave}[mmWave]{millimeter wave}
\acrodef{mMIMO}[mMIMO]{massive multiple-input multiple-output}
\acrodef{AP}[AP]{access point}
\acrodef{6-G}[6-G]{sixth-generation}

\maketitle
\begin{abstract}
We consider a cell-free \ac{mMIMO} network, where \acp{UAV} equipped with multiple antennas serve as distributed \acp{UAV-AP}. These \acp{UAV-AP} provide seamless coverage by jointly serving \acp{UE} without predefined cell boundaries. However, high-capacity wireless networks face significant challenges due to fronthaul limitations in \ac{UAV}-assisted architectures. This letter proposes a novel \ac{UAV}-based cell-free \ac{mMIMO} framework that leverages distributed UAV-APs to serve \acp{UE} while addressing the capacity constraints of wireless fronthaul links. We evaluate functional split Options 7.2 and 8 for the fronthaul links, aiming to maximize the minimum \ac{SINR} among the \acp{UE} and minimize the power consumption by optimizing the transmit powers of \acp{UAV-AP} and selectively activating them. Our analysis compares sub-6 GHz and \ac{mmWave} bands for the fronthaul, showing that mmWave achieves superior SINR with lower power consumption, particularly under Option 8. Additionally, we determine the minimum fronthaul bandwidth required to activate a single UAV-AP under different split options.

\end{abstract}


\section{Introduction}

The integration of \acp{UAV} into \ac{6G} technology is poised to revolutionize multiple industries by significantly enhancing UAV capabilities. Leveraging the ultra-low latency of 6G, \acp{UAV-AP} will be able to transmit high-speed data and support real-time applications such as remote surveillance, high-resolution 3D mapping, and augmented reality~\cite{9}. However, \acp{UAV-AP} typically rely on wireless links for communication with ground stations and other nodes, and the bandwidth available for fronthaul connections is often limited. Additionally, wireless data transmission consumes substantial energy, underscoring the importance of power optimization to extend flight times and operational ranges. In addition, in dense and highly dynamic environments, high levels of interference present significant challenges for traditional cell-based systems in delivering adequate coverage and capacity.

To address these issues, cell-free \ac{mMIMO} has emerged as a transformative technology for next-generation wireless networks. By leveraging joint transmission and reception from a large number of distributed \acp{AP}, cell-free \ac{mMIMO} significantly enhances spectral efficiency~\cite{4}. However, the distributed architecture and joint processing requirements of cell-free \ac{mMIMO} demand efficient resource allocation both in the access and the fronthaul. In \cite{5}, the authors propose end-to-end network power consumption minimization by considering intra-physical layer (PHY) functional splits, focusing on Options 7.2 and 8. Similarly, \cite{15} investigates uplink cell-free UAV networks across different fronthaul transport strategies such as \ac{FDMA} and \ac{SDMA}. The study optimizes the UAV deployment and UAV transmit powers to maximize the minimum spectral efficiency. Despite these advances, \acp{UAV} are constrained by limited battery capacity, making power consumption a critical issue in UAV-assisted networks. This constraint underscores the need for carefully selecting functional split options to optimize resource allocation. Lower-layer splits, such as Option 8, are often preferred for UAVs as they confine their role to \ac{RF} transmission/reception, thereby reducing on-board processing demands. Conversely, higher-layer splits, like Option 7.2, are better suited for wireless fronthaul links, as they offload much of the processing to the \acp{UAV-AP} and reduce signaling needs, helping to address fronthaul capacity constraints. While existing studies have explored power minimization and UAV placement strategies, they have not thoroughly examined the impact of limited fronthaul capacity constraints, particularly when considering functional split options to jointly optimize the UAV placement and power allocation.

Motivated by this, this letter explores two functional split options for the wireless fronthaul between \ac{CPU} and \acp{UAV-AP}, where \acp{UAV} serve ground \acp{UE} while ensuring a reliable and efficient wireless fronthaul connection. The main contributions of the letter are as follows: 1) We develop a novel model for UAV-assisted cell-free mMIMO network, specifically designed to enhance UE access while satisfying the stringent capacity constraints of fronthaul links. 2) By considering fronthaul constraints, we minimize the power consumption of UAV-APs by optimizing their transmit powers while ensuring a fair SINR requirement is met for all \acp{UE}. 
3) We compare sub-6 GHz and \ac{mmWave} bands in wireless fronthaul links to study their impact on SINR and UAV power consumption.
4) We analyze the trade-off between functional split Options 7.2 and 8 at the wireless fronthaul, focusing on their impact on SINR experienced by the \acp{UE} and the power consumption at the \acp{UAV-AP}.

\section{System Model}
We consider a downlink cell-free \ac{mMIMO} system consisting of $L$ \acp{UAV-AP} and $K$ \acp{UE} on the ground with orthogonal frequency division multiplexing (OFDM).  Each \ac{UAV-AP} is equipped with $N_a$ antennas and serves single-antenna \acp{UE} in the access link. 
% The system is built on top of a next-generation virtualized \ac{C-RAN} architecture, which provides fronthaul connectivity to the \acp{UAV-AP}. 
Each \ac{UAV-AP} is connected to the \ac{CPU}, located at the center of the coverage area at a specific height, and equipped with $N_c$ antennas, via a single fronthaul antenna. Due to the wireless fronthaul capacity limitations, we compare two low PHY functional split options, Option 8 and Option 7.2~\cite{12}. The \ac{3GPP} protocol stack is divided between two places: the \ac{RU} and the \ac{DU}. In Option 8, only the \ac{RF} sampler and upconverter are placed in \acp{RU}, which are the \acp{UAV-AP}, and the remaining functions are centralized. 
% The baseband signals that are sampled and quantized are received from the \ac{CPU}, and the \ac{RF} processing is performed at the \acp{UAV-AP}, which acts as remote radio heads (RRH). 
Therefore, the downlink fronthaul rate requirement for each \ac{UAV-AP} for Option 8 is given as~\cite{6},\cite{5}
\begin{equation}
    R^{f(8)}= 2 f_sN_{\mathrm{bits}} N_a
    \label{op8r}
\end{equation}
where $f_s$ is the sampling frequency and $N_{\mathrm{bits}}$ is the number of bits required to quantize the signal samples, and the factor 2 refers to the complex nature of signals. In Option 7.2, the low-PHY functions, such as discrete Fourier transform (DFT) operations, are included locally in the UAV-APs, and the high-PHY functions and the remaining higher-layer operations are centralized. 
% Here, UAV-APs act as macro BSs. 
Therefore, the downlink fronthaul rate requirement for each UAV-AP for Option 7.2 is given as~\cite{6},\cite{5}
\begin{equation}
    R^{f(7.2)}= \frac{2 N_{\mathrm{bits}} N_{\mathrm{used}} N_a}{T_s}
    \label{op72r}
\end{equation}
where $N_{\mathrm{used}}$ is the number of used subcarriers neglecting those subcarriers used as a guard band, and $T_s$ is the OFDM symbol duration. Typically, $R^{f(8)} \textgreater R^{f(7.2)}$, since $f_s \textgreater \frac{N_{\textrm{used}}}{T_s}$. 

\subsection{Wireless Fronthaul}

We consider a \ac{MU-MIMO} setup where a \ac{CPU} equipped with $N_c$ antennas communicates with UAV-APs, each having a single antenna dedicated to the fronthaul, over a shared fronthaul channel with a total bandwidth of $B_F$ Hz. A uniform planar array with half wavelength antenna spacing is considered at the CPU.
% We use \ac{SDMA} due to the fronthaul capacity limitations. 
The received signal $\mathbf{y}_f=[y_1,y_2,...,y_L]^{\mathrm{T}}$ for all the \acp{UAV-AP} are given as
\begin{equation}
\mathbf{y}_f= \mathbf{H}_f^{\mathrm{T}} \mathbf{P} \mathbf{x} + \mathbf{n},
\end{equation}
where $\mathbf{H}_f= [\mathbf{h}^f_1, \mathbf{h}^f_2,...,\mathbf{h}^f_L]$ is the channel matrix where $\mathbf{h}_l^f \in \mathbb{C}^{N_c}$ denote the channel between the \ac{CPU} and UAV-AP $l$ in the fronthaul link. The receiver noise $\mathbf{n}$ has independent $ \mathcal{N}_{\mathbb{C}}(0,B_FN_0)$ entries, where  $N_0$ is the noise power spectral density. The maximum transmit power at the CPU to be allocated to the $L$ UAV-APs is given as $P_{\rm{max}}^f$. We assume perfect \ac{CSI} is available at the fronthaul link for analytical tractability. We consider a zero-forcing precoding matrix $\mathbf{P}^{\mathrm{ZF}}$ so that the interference at co-UAV-APs are nullified~\cite[Ch.~6]{3}: 
\begin{equation}
    \mathbf{P}^{\mathrm{ZF}}= \mathbf{H}_f^*\left(\mathbf{H}_f^{\mathrm{T}} \mathbf{H}_f^*\right)^{-1} \mathbf{Z}^{\mathrm{ZF}}
        \label{zfpr}
\end{equation} 
where
\begin{equation}
   \mathbf{Z}^{\mathrm{ZF}}\!\!\!\!=\!\mathrm{diag}\!\left(1/\sqrt{[(\mathbf{H}_f^{\mathrm{T}} \mathbf{H}_f^*)^{-1}]_{11}},\ldots,1/\sqrt{[(\mathbf{H}_f^{\mathrm{T}}  \mathbf{H}_f^*)^{-1}]_{LL}}\right).
\end{equation}

The expression for the channel matrix depends on the frequency band. Here, $\mathbf{H}{^\mathrm{T}}$ represents the transpose of a matrix $\mathbf{H}$, and $\mathbf{H}^*$ represents its conjugate. We compare sub-6 GHz and \ac{mmWave} bands in the wireless fronthaul to analyze their performance differences. The fronthaul bandwidth $B_F$ and the fronthaul antennas $N_c$ between \ac{CPU} and \acp{UAV-AP} vary depending on the frequency band.

\subsubsection{\ac{mmWave} channel}
An extended Saleh-Valenzuela channel model is considered with one direct path and a few scattered paths between the CPU and the UAV-APs. The channel of UAV-AP $l$ is given as 
\begin{equation}
    \mathbf{h}_l^f= \sum_{i=0}^n \alpha_i  \mathbf{a}_{i}(\varphi_{l,i}^f,\theta_{l,i}^f)^{\mathrm{T}}
    \label{ric_fad}
\end{equation}
where $n$ denotes the number of scattered paths, where $i=0$  corresponds to the direct path. $\mathbf{a}_{i}(\varphi_l^f,\theta_l^f)$ is the array response vector given as in \cite{10}. $\alpha_i$ is the complex gain of the direct and scattered paths. For the scattered paths, we have $\alpha_i \thicksim \mathcal{N}_{\mathbb{C}}(0, \beta_i)$, where $\beta_i$ is the corresponding average power. 

\subsubsection{sub-6 GHz}
The channel is expressed as $\mathbf{h}^f_l= \bar{\mathbf{h}}^f_l +\Tilde{\mathbf{h}}^f_l
$ where $\bar{\mathbf{h}}^f_l$ is the \ac{LoS} component~\cite{11} for which the potential random phase-shift is included in the corresponding array response vector.  $\Tilde{\mathbf{h}}^f_l$ represents the \ac{NLoS} components. 
% is given as 
% % $\bar{\mathbf{h}}^f_l= e^{-j\phi_l^f}\sqrt{\bar{\beta}_l}\mathbf{a}_t^{l}(\varphi_l^f,\theta_l^f)$, 
% $\bar{\mathbf{h}}^f_l= \sqrt{\bar{\beta}_l^f}\mathbf{a}_t^{l}(\varphi_l^f,\theta_l^f)$, 
% where $\bar{\beta}_l^f$ is the LOS channel gain~\cite{2} and the array response vector at the CPU is given in \cite{10}. 
Compared to mmWave channels, the sub-6 GHz channels are expected to have a richer scattering nature. Accordingly, the \ac{NLoS} channels $\Tilde{\mathbf{h}}^f_l$ are modeled as correlated Rayleigh channels with a spatial correlation matrix generated according to the local scattering model \cite{4}.

Considering either of the channel models, we apply the ZF precoding given in \eqref{zfpr} to suppress the interference at the fronthaul link. The downlink rate experienced by UAV-AP $l$ is then given as
$ R^f_{l}= B_F \log_2\bigg(1+\frac{P^f_{l}}{B_F N_0 \big[(\mathbf{H}_f^{\mathrm{T}}  \mathbf{H}_f^*)^{-1}\big]_{ll}}\bigg)$
where $P^f_{l}$ is the transmit power at the CPU allocated to  UAV-AP $l$. The transmit power constraint at the CPU is $  \sum_{l=1}^L P_{l}^{f } \leq P_{\mathrm{max}}^f$. 
Considering the fronthaul data rate requirement for Option 8 and Option 7.2 given in \eqref{op8r} and \eqref{op72r}, respectively, we can find the minimum transmit power to be allocated to each UAV-AP from the CPU. Considering $\kappa \in \{8,7.2\}$, and rearranging the above equation, the transmit power allocated to UAV-AP $l$ is 
\begin{equation}
  P_{l}^{f(\kappa)}= \left(2^{\frac{R_{l}^{f (\kappa)}}{B_F}}-1\right) B_F N_0  [(\mathbf{H}_f^{\mathrm{T}}\mathbf{H}_f^*)^{-1}]_{ll},  
  \label{fh_limit}
\end{equation}
\subsection{Cell-free Access}
We consider a cell-free \ac{mMIMO} setup employing a time-division duplex protocol, where the UAV-APs are equipped with $N_a$ antennas transmitting to single-antenna UEs. The system utilizes OFDM, dividing the available bandwidth $B_A$ into numerous subcarriers. %The carrier frequency and sampling frequency are represented as $f_c$ and $f_s$, respectively. 
% A block-fading channel model is considered~\cite{4}, where time-frequency resources are divided into coherence blocks. % Within each block, the channel remains time-invariant and frequency-flat, accommodating several sub-carriers.
Without loss of generality, we consider the channel to be constant over $\tau_c$ time-frequency samples in each coherence block. We consider distributed cell-free operation, in which the channel estimation is performed at the UAV-APs for the local precoding during the data transmission. Each coherence block $\tau_c$ is divided into two parts: $\tau_u$ samples for uplink channel estimation and $\tau_d$ samples for data transmission. We use the sub-6 GHz band in the access link. Because the channel conditions fluctuate over time due to fading, it becomes necessary to estimate the channel parameters for each coherence block. 
% We consider the same channel modeling as the sub-6 GHz band in the fronthaul, with an additional phase shift of $\phi_l^{a}$ in the LoS component.
% We assume imperfect \ac{CSI} is available at the access link. 
Therefore, we apply the linear minimum mean square error (LMMSE) channel estimation
\cite{4}, \cite{5}. We consider the L-MMSE precoding technique \cite{5} to construct a good balance between maximizing the signal strength and canceling the interference. 
% For a UAV-AP $l$ that serves user $k$, the precoding vectors are considered the same as \cite{5}. 
% $
% \mathbf{w}_{lk}=\frac{\bar{\mathbf{w}}_{lk}}{\sqrt{\mathbb{E}\{\|\bar{\mathbf{w}}_{lk}\|^2\}}}$. The error correlation matrix $\mathbf{M}_{lk}= \mathbb{E}\{(\mathbf{h}_{lk}-\hat{\mathbf{h}}_{lk})(\mathbf{h}_{lk}-\hat{\mathbf{h}}_{lk})^{\mathrm{H}}\}$.
% where $\rho_{lk}$ is the transmit power from UAV-AP $l$ to user $k$. $\bar{\mathbf{w}}_{lk}$ is the arbitrary scale vector pointing out the direction of the precoding vector $\mathbf{w}_{lk}$~\cite{4}. Using L-MMSE precoding, 
% \begin{equation}
% \bar{\mathbf{w}}_{lk}=\eta_k \Bigg( \sum_{i=1}^{K} \eta_i \Big(\hat{\mathbf{h}}_{li}(\hat{\mathbf{h}}_{li})^{\mathrm{H}} + \mathbf{M}_{li} \Big) + \sigma_a^2 \mathbf{I}_{N_a} \Bigg)^{-1} \hat{\mathbf{h}}_{lk}
% \end{equation}

%We let $\mathcal{M}_k$ be the set of indices of the UAV-APs which are serving user $k$. 
We let $\mathbf{h}_{kl}$ and $\mathbf{w}_{kl}$ denote the channel and  normalized (in terms of average power)
precoding vector  of UE $k$ at UAV-AP $l$, respectively. We also denote the power assigned to UE $k$ by UAV-AP $l$ by $P_{kl}^a$. The spectral efficiency experienced by UE $k$ is given as
$\mathcal{S}_k= \frac{\tau_d}{\tau_c} \log_2(1+\gamma_k)$, where the effective downlink \ac{SINR} experienced by UE $k$ is given as~\cite{4}
\begin{equation}
    \gamma_k (\{\boldsymbol{\rho}_i\})= \frac{|\mathbf{b}_k^{\mathrm{T}} \boldsymbol{\rho}_k|^2}{\sum_{i=1}^K \boldsymbol{\rho}_i^{\mathrm{T}} \mathbf{C}_{ki} \boldsymbol{\rho}_i-|\mathbf{b}_k^{\mathrm{T}} \boldsymbol{\rho}_k|^2 + \sigma_a^2},
    \label{sinr_1}
\end{equation}
where, $\sigma^2_a$ is the noise power and
 \begin{align}   
 & \boldsymbol{\rho}_k=[\sqrt{P_{k1}^a} ,\ldots,\sqrt{P_{kL}^a} ]^{\mathrm{T}}= [\rho_{k,1} ,\ldots,\rho_{k,L} ]^{\mathrm{T}}, &
 \label{rho_k}\\
     & \mathbf{b}_k \in \mathbb{R}_{\geq 0}^L, [\mathbf{b}_k]_l=\mathbb{E}\{\mathbf{h}_{kl}^{\mathrm{T}} \mathbf{{w}}_{kl}\}, \\
     &\left[ \mathbf{C}_{ki} \right]_{lr} = 
    \mathbb{E} \left\{ \mathbf{h}_{kl}^{\mathrm{T}}\mathbf{w}_{il} \mathbf{w}_{ir}^{\mathrm{H}} \mathbf{h}_{kr}^* \right\}.
\end{align}
\section{Max-Min Fairness}
Max-min fairness in a network is to maximize the minimum downlink effective \ac{SINR} among all the UEs on the ground. Here, we consider the total fronthaul transmit power $P_{\rm max}^f$, and the maximum allowed transmit power of UAV-APs $P_{\rm UAV}$ as constraints. The aim is to optimize the transmit power allocated to the UEs, i.e., $P_{kl}^a$, and to decide which UAV-APs are active and serving the UEs, where $\boldsymbol{\alpha}  = [\alpha_1, \ldots, \alpha_L]^{\mathrm{T}} \in \mathbb{B}^{L}$ denotes the binary UAV-AP activation vector. $\alpha_l=1$ indicates that UAV-AP $l$ is active and $\alpha_l=0$ means inactive.
To express the constraints and the objective function as a mixed binary linear or conic form, the power coefficients are represented as
$\boldsymbol{\rho}_k$ given in (\ref{rho_k}) and we define $\overline{\boldsymbol{\rho}}_l=[\rho_{1,l} , \ldots,\rho_{K,l}]^{\mathrm{T}}$ with all the powers related to the UAV-AP $l$.

The optimization problem can be cast as
\begin{subequations} \label{ver_1}
\begin{align} 
\underset{\boldsymbol{\rho}_{k},\alpha_l \forall k,l}{\textrm{maximize}} \quad & \min_{k=1, \dots, K} \quad \gamma_k (\{\boldsymbol{\rho}_i\}) \label{main_1}
% \frac{|\mathbf{b}_k^{\mathrm{T}} \boldsymbol{\rho}_k|^2}{\sum_{i=1}^K \boldsymbol{\rho}_i^{\mathrm{T}} \mathbf{C}_{ki} \boldsymbol{\rho}_i + \sigma_a^2} 
\\
\textrm{subject to} \quad 
& \sum_{l=1}^L \alpha_l P^f_{l} \leq P_{\mathrm{max}}^f , \tag{\ref{ver_1}b} \label{constr1} \\
& ||\overline{\boldsymbol{\rho}}_l|| \leq \alpha_l \sqrt{P_{\mathrm{UAV}}}, \quad \forall l , \tag{\ref{ver_1}c} \label{constr2} \\
&\alpha_l \in \{0,1\}, \quad \forall l.
\tag{\ref{ver_1}d} \label{constr3}
\end{align} 
\end{subequations}
The constraint in \eqref{constr1} ensures that the total fronthaul power assigned by the CPU to the active UAV-APs should be less than the total power allowable at the fronthaul. Accordingly, the UAV-APs are disabled or enabled such that this constraint is satisfied. \eqref{constr2} is the constraint on the maximum transmit power per UAV-AP, and \eqref{constr3} represents the binary UAV-AP selection variables.

This optimization problem can be solved using the bi-section search algorithm over $t^c$ that corresponds to the minimum of the UE SINRs. For faster convergence, we replace the feasibility check problem with a total power minimization problem. Therefore, the optimization problem changes to a second-order cone problem with binary variables, where the following optimization problem is solved at each iteration.
\begin{subequations}
\begin{align}
    \underset{\boldsymbol{\rho}_{k},\alpha_l \forall k,l}{\textrm{minimize}} \quad & \sum_{k=1}^K \|\boldsymbol{\rho}_k\|^2 \\
    \textrm{subject to} &\quad \eqref{constr1}-\eqref{constr3}, \\
    & \left\|\begin{bmatrix}
    \mathbf{C}_{k1}^{\frac{1}{2}}\boldsymbol{\rho}_1 \\
    \vdots \\ \mathbf{C}_{kK}^{\frac{1}{2}}\boldsymbol{\rho}_K \\
    \sigma_a \end{bmatrix}\right\| 
    \leq \sqrt{\frac{1 + t^c}{t^c}} \mathbf{b}_k^{\mathrm{T}} \boldsymbol{\rho}_k, \quad \forall k. 
    % & \sum_{l=1}^L \alpha_l P^f_{l} \leq P_{max}^f \\
    % & \|\overline{\boldsymbol{\rho}}_l\| \leq \alpha_l \sqrt{P_{UAV}} \quad l = 1, \dots, L\\
    % & \alpha_L \in \{0,1\} \quad \forall L
\end{align}
\end{subequations}
The solution is calculated over each value of $t^c$~\cite{4}. The value of $t_c$ is updated based on the problem's feasibility, continuing until the largest feasible point is reached. By substituting the optimal values of transmit powers into \eqref{sinr_1} and taking a minimum SINR, we obtain the maximized minimum SINR of the UEs given as $\gamma_{\mathrm{opt}}=\min\limits_{k \in \{1,\dots, K\}} \gamma_k(\{\boldsymbol{\rho}_i^{\mathrm{opt}}\})$. 

\section{Total UAV-AP Power  Minimization}

In this section, we discuss the total power consumption at the UAV-APs, considering the hardware limitations of fixed-wing \acp{UAV-AP} and processing powers. Fixed-wing UAVs are designed for long-range missions and have high service time. The total power consumption consists of the total power required to establish a reliable fronthaul link between the CPU and UAV-APs. Therefore, the total power consumption in a UAV-based network consists of 
% i) mechanical power consumption
i) fronthaul power consumption as given in (\ref{fh_limit}); ii) RF transmission power consumption; iii) processing power for Option 7.2. Option 8 does not consume processing power, only RF transmission power.

% \subsection{Mechanical power consumption}

% In this work, we consider a rotary-wing drone to serve the users on the ground establishing a fronthaul to the CPU. The mechanical power consumption for rotary-wing drones in hovering configuration is given as
% \begin{equation}
%  P_{RW}=\frac{\delta}{8} \mu \psi A \Omega^3 R^3 +(1+\vartheta) \frac{W^3/2}{\sqrt{2 \mu A}},   
% \end{equation}
% where $\delta$ is the profile drag coefficient, $\mu$ is the air density, $\psi$ is the rotor solidity, $A$ is the rotor disc area, $\Omega$ is the blade angular velocity, $R$ is the rotor radius, $\vartheta$ is the incremental correction factor and $W$ is the weight of rotary-wing drone. 
% \begin{itemize}
\textbf{RF transmission power consumption:} The RF transmission power consumption consists of the transmit power of a single UAV-AP, which is given as $||\overline{\boldsymbol{\rho}}_l||^2$.  The load-dependent power consumption is modeled by the slope $\Psi_t$. The total power consumption consists of the total number of UAV-APs selected. There is static power consumption for each UAV-AP when no transmission occurs, which is combined in this term. 

\textbf{Processing power consumption:} The baseband processing is performed at the UAV-APs for Option 7.2. Based on a load-dependent power consumption model, $P_{{\rm UAV},0}^{\mathrm{proc}} + \frac{\Psi_{{\rm UAV}}^{\mathrm{proc}} \mathcal{P}}{C_{{\rm UAV}}^{\mathrm{max}}}$ gives the total processing power consumption. Here, $P_{{\rm UAV},0}^{\mathrm{proc}}$ is the processing power of each UAV-AP in idle mode, $\Psi_{{\rm UAV}}^{\mathrm{proc}}$ is the slope of the load-dependent processing power consumption, $C_{{\rm UAV}}^{\mathrm{max}}$ is the maximum processing capacity of each UAV-AP given in \ac{GOPS}, and $\mathcal{P}$ is the total GOPS in one UAV-AP. After the RF operations, polyphase baseband filtering is performed. The complexity to perform this filtering operation per UAV-AP is given as $C_{F}= \frac{40 N_a f_s}{10^9}$ GOPS~\cite{5}. After filtering, inverse DFT is performed with a complexity of    $C_{D}=\frac{8 N_a N_{\mathrm{DFT}} \log_2(N_{\mathrm{DFT}})}{T_s 10^9}$ GOPS~\cite{5}. $N_{\mathrm{DFT}}$ is the total number of subcarriers and the dimension of the DFT. Therefore, the total GOPS in one UAV-AP for processing is 
$\mathcal{P}= C_D+ C_F$.
% \end{itemize}

The input variable $\mathbb{I} \in \{0,1\}$ indicates which functional split option is selected. If Option 8 is selected, $\mathbb{I}=0$ and for Option 7.2, $\mathbb{I} =1$.
    The total power consumption at the UAV-AP is 
    \begin{multline}
    \rho_{\mathrm{tot}}(\{\boldsymbol{\rho}_i,\alpha_l\})= \mathbb{I}\bigg(P_{{\rm UAV},0}^{\mathrm{proc}} + \frac{\Psi_{{\rm UAV}}^{\mathrm{proc}} \mathcal{P}}{C_{{\rm UAV}}^{\mathrm{max}}}\bigg) \sum_{l=1}^L \alpha_l+  \\ \sum_{l=1}^L \alpha_l \bigg(
    \mathbb{I}P_{l}^{f(7.2)} +(1-\mathbb{I})P_{l}^{f(8)}\bigg) + \sum_{l=1}^L \alpha_l  \cdot P_{A}+ \Psi_t \sum_{l=1}^L \sum_{k=1}^K \rho_{k,l}^2,
    \end{multline}
    where $P_{A}$ is the power consumption due to the power amplifier. A factor two of overhead is included in $P_{A}$ to account for memory operations. We consider minimization of total power consumption in UAV-APs, assuming an SINR requirement of $\Gamma_k$ for each UE in the access link and the fronthaul power limitations. The optimization problem is formulated as
\begin{subequations} 
\begin{equation}
  \underset{\boldsymbol{\rho}_{k},\alpha_l \forall k,l}{\textrm{minimize}} \quad  \rho_{\mathrm{tot}}(\{\boldsymbol{\rho}_i,\alpha_l\})  
\end{equation}
\begin{align}
\textrm{subject to} \quad & 
% \frac{|\mathbf{b}_k^{\mathrm{T}} \boldsymbol{\rho}_k|^2}{\sum_{i=1}^K \boldsymbol{\rho}_i^{\mathrm{T}} \mathbf{C}_{ki} \boldsymbol{\rho}_i + \sigma_a^2} \geq \Gamma_k
\left\|\begin{bmatrix}
    \mathbf{C}_{k1}^{\frac{1}{2}}\boldsymbol{\rho}_1 \\
    \vdots \\ \mathbf{C}_{kK}^{\frac{1}{2}}\boldsymbol{\rho}_K \\
    \sigma_a \end{bmatrix}\right\|
\leq \sqrt{\frac{1 + \Gamma_k}{\Gamma_k}} \mathbf{b}_k^{\mathrm{T}} \boldsymbol{\rho}_k, \quad \forall k ,
\label{const_1}\\
% \left\|\begin{bmatrix}
% \tilde{\mathbf{C}}_{k1}^{\frac{1}{2}}\tilde{\rho}_1 \\
%     \vdots \\ \tilde{\mathbf{C}}_{kK}^{\frac{1}{2}}\tilde{\rho}_K \\
% \sqrt{\sigma^2_{dl}}
% \end{bmatrix}\right\| 
% \leq \sqrt{\frac{1 + t}{t}} \tilde{\mathbf{b}_k^T} \tilde{\rho}_k, \quad k = 1, \dots, K \\
& \sum_{l=1}^L \alpha_l \Big(
\mathbb{I}P_{l}^{f(7.2)} +(1-\mathbb{I})P_{l}^{f(8)}\Big) \leq P_{\mathrm{max}}^f ,
\label{const_2}\\
& ||\overline{\boldsymbol{\rho}}_l|| \leq \alpha_l \sqrt{P_{\mathrm{UAV}}}, \quad \forall l,
\label{const_3}\\
& \alpha_l \in \{0,1\}, \quad \forall l.
\label{const_4}
\end{align}
\end{subequations}
The constraint in \eqref{const_1} satisfies the minimum SINR requirement for each UE, $\Gamma_k$, represented as a second-order cone constraint. The constraint \eqref{const_2} limits the allocated fronthaul transmit powers for each UAV-AP according to the maximum allowable fronthaul power $P_{\mathrm{max}}^f$. The constraint in \eqref{const_3} provides the transmit power constraint per UAV-AP. \eqref{const_4} defines $\alpha_l$ as the binary activation variable.

This is a mixed binary second-order cone problem; thus, it is non-convex and combinatorial due to the binary variables. However, it can be solved using any mixed binary second-order cone programming solver. We solved this problem with the CVX-MOSEK solver~\cite{16}. %By substituting the values of the variables, we can determine the minimized total power consumption of the UAV-APs while serving the users on the ground, all while adhering to the power constraints imposed on the fronthaul. 
\begin{figure*}
    \centering

    \subfigure[\label{bar_FH}]{\includegraphics[width=0.24\linewidth, height=0.2\linewidth]{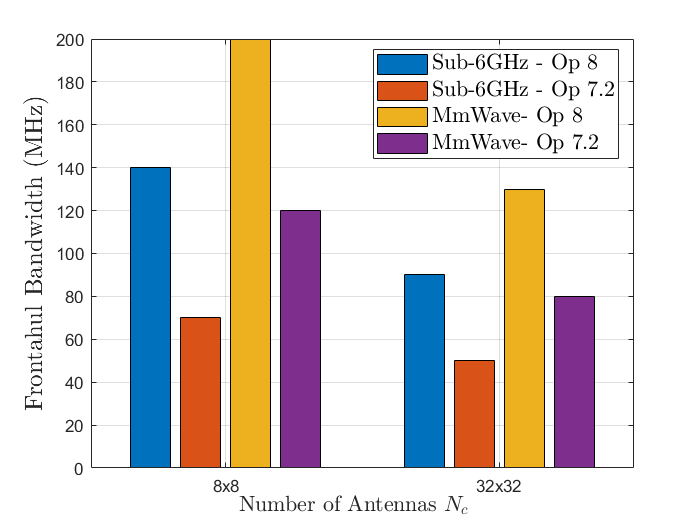}}
    \subfigure[\label{SINR_Na}]{\includegraphics[width=0.24\linewidth, height=0.2\linewidth]{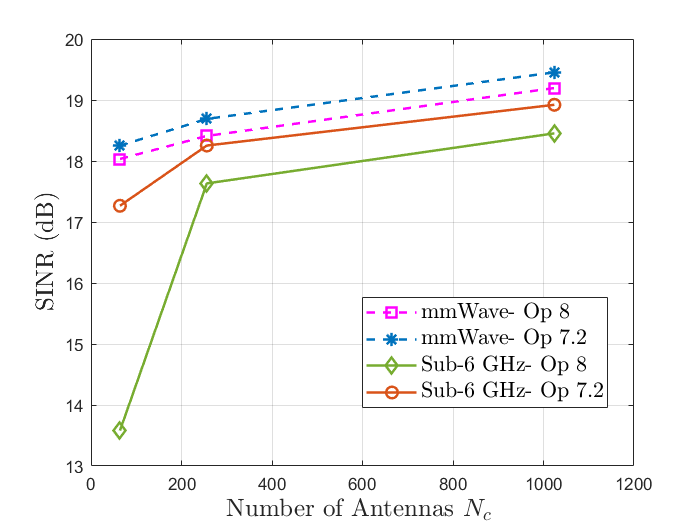}}
    \subfigure[\label{PC_REQ}]{\includegraphics[width=0.24\linewidth, height=0.2\linewidth]{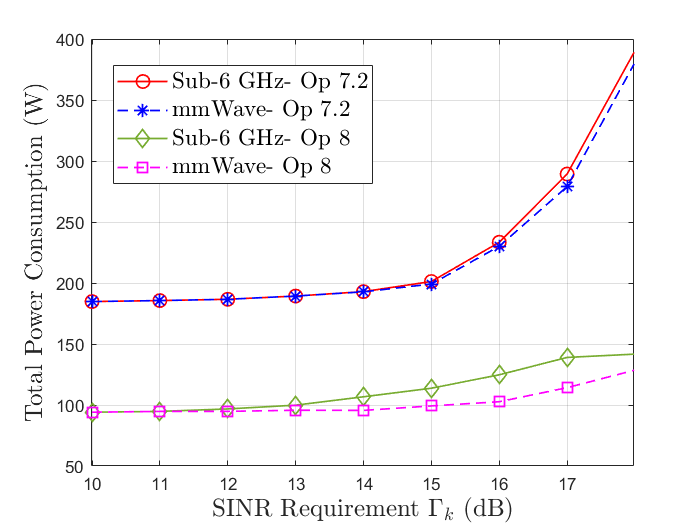}}
    \subfigure[\label{bar_op}]{\includegraphics[width=0.24\linewidth, height=0.2\linewidth]{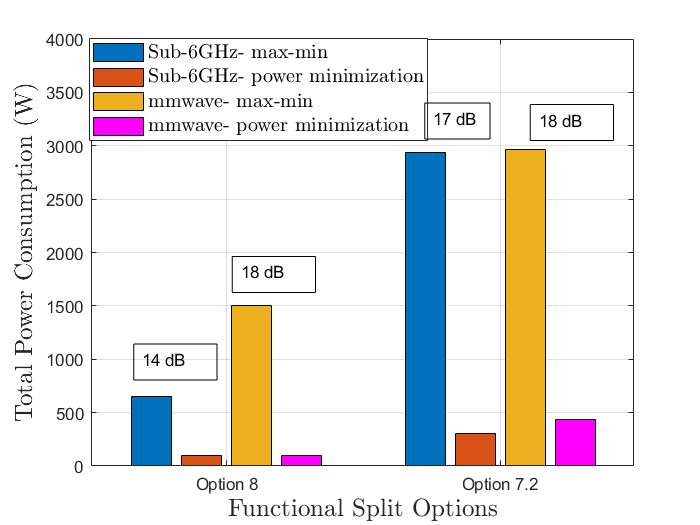}}
    \caption{a) Number of fronthaul antennas versus the fronthaul bandwidth to activate one UAV-AP for Options 8 and 7.2. b) Number of fronthaul antennas versus the SINR at the UE  with fronthaul bandwidth $B_F=500$\,MHz for \ac{mmWave}, $B_F=150$\,MHz for sub-6 GHz for Options 8 and 7.2. c) Total power consumption versus SINR requirement with $B_F=500$\,MHz, $N_c= 1024$ for \ac{mmWave}, $B_F=150$\,MHz, $N_c= 64$ for sub-6 GHz, for Options 8 and 7.2. d) Comparison for max-min fairness optimization and power minimization for different options with maximized SINR requirement.}
\end{figure*}

\section{Numerical Results and Discussion}

In this section, we assess the downlink SINR experienced by the UEs and the total power consumption in a UAV-based network, taking into account the capacity limitations of the wireless fronthaul link. We compare two functional split options, analyzing their impact on the UEs' \ac{QoS} and overall power consumption at the UAV-APs. There are $L=16$ UAV-APs and $K=8$ UEs. We set $N_a=4$, $P_{\mathrm{max}}^f= 10 \, \mathrm{W}$, $P_{\mathrm{UAV}}=1 \, \mathrm{W}$, and the pilot power as $100$\,mW.
The sub-6 GHz-band system simulation parameters are $N_c=64$, $f_c= 3.5 \, \mathrm{GHz}$. The \ac{mmWave} simulation parameters are $N_c=1024$, $f_c= 28 \, \mathrm{GHz}$, and the number of scattered paths is $n=1$ for the fronthaul link. The RRH power and processing configurations are considered from \cite{7}, \cite{13}, \cite{14}. The processing power parameters are $P_{{\rm UAV},0}^{\mathrm{proc}}= 20.8 \, \mathrm{W}$, $\Psi_{{\rm UAV}}^{\mathrm{proc}}= 74\, \mathrm{W}$ and $C_{{\rm UAV}}^{\mathrm{max}}= 180 \ \mathrm{GOPS}$~\cite{5}. The other parameters are $N_{\mathrm{DFT}}= 2048$, $N_{\mathrm{used}}= 1200$, $f_s= 30.72 \, \mathrm{MHz}$, $T_s=71.4 \, \mathrm{\mu s}$, $\tau_c= 192$, $\tau_u= 8$, $N_{\mathrm{bits}}=8$ and $P_A= 64.4 \ \mathrm{W}$~\cite{8}. 
% The mechanical power consumption parameters~\cite{8} are $\delta=0.012$, $\mu= 1.225 \mathrm{Kg/m^3}$,$\psi=0.05$, $R=0.4 \mathrm{m}$, $\Omega= 300 \mathrm{rad/s}$, $\vartheta= 0.1$, $A_D= 0.503$, $W= 2.03$.
The UEs and the UAV-APs are distributed uniformly in an area of 1 km $\times$ 1 km. We consider fixed-wing drones at a height $h_u= 200 \, \mathrm{m}$ to serve the UEs on the ground, establishing a fronthaul link to the CPU at the height of $h_c =50 \, \mathrm{m}$.  
% The channel between the CPU and UAV-AP $l$ is represented as 
% $\Tilde{\mathbf{h}}^f_l \thicksim \mathcal{N}_{\mathbb{C}}(\mathbf{0}_{N_c},\Tilde{\mathbf{R}}^f_l)$
% where $\Tilde{\mathbf{R}}^f_l \in \mathbb{C}^{N_c \times N_c}$ is the spatial correlation of the channel $\Tilde{\mathbf{h}}^f_l$ between the $N_c$ antenna of CPU connecting to the UAV-AP $l$. From the diagonal elements of $\Tilde{\mathbf{R}}^f_l$, we define the average channel gain between an antenna at CPU and the UAV-AP $l$. The average channel gain $\beta_l^f$= $\frac{1}{N_c} \mathrm{tr}(\Tilde{\mathbf{R}}^f_l)$, which depends on the large-scale effects such as path loss, geometric attenuation, and shadowing~\cite{2}. 
We follow the same path loss model as \cite{1} for the fronthaul and access links, with the probability of \ac{LoS} is defined as
    % \begin{equation}
    $P_L=\frac{1}{1+\eta_1 \exp{(-\eta_2(\theta_l-\eta_1))}}$
    % \label{los_p}
% \end{equation}
where $\eta_1$ and $\eta_2$ are the environmental parameters and $\theta_l$ is the elevation angle. The environmental parameters in the fronthaul are $\eta_1^f= 4.8$, $\eta_2^f= 0.43$, and in the access link are $\eta_1^a= 9.61$ and $\eta_2^a= 0.16$. The mechanical power consumption of fixed-wing drones, as provided in~\cite{8}, is not considered in the optimization problem but is incorporated into the calculation of the service time of UAV-APs.

In Fig.~\ref{bar_FH}, we plot the minimum fronthaul bandwidth required to activate a single UAV-AP in the network with the best fronthaul channel to provide access to the UEs on the ground. We observe a significant reduction in fronthaul bandwidth requirement in Option 7.2 compared to Option 8. Because of the lower rate requirement, the fronthaul bandwidth requirement in Option 7.2 is reduced by around 55\% compared to Option 8. The mmWave channels required 42\% more bandwidth because of the less favorable propagation conditions, but the required bandwidth is feasible for real-time applications, and these bands offer a larger available spectrum. 

In Fig.~\ref{SINR_Na}, we plot the number of antennas at the CPU $N_c$ versus the \ac{SINR} experienced by the UEs using the max-min fairness algorithm. We observe that Option 7.2 is performing better than Option 8 as it allows a higher number of UAV-APs to be active because of the lower fronthaul requirement, leading to an increase in \ac{SINR} in the network. Moreover, we observe that mmWave channels achieve higher \ac{SINR} compared to sub-6 GHz, attributed to their capacity to support the activation of more UAV-APs. However, the SINRs obtained for Option 8 and Option 7.2 are very close in the mmWave channel. 

In Fig.~\ref{PC_REQ}, we plot the total power consumption at the UAV-APs in terms of the \ac{SINR} requirement at the UE end. We observe that Option 7.2 consumes 90\% more power than Option 8 because of the processing functions performed at the UAV-APs, which leads to more power consumption at UAV-APs. We observe that mmWave consumes slightly less power due to its higher fronthaul capacity, enabling it to achieve the same SINR as sub-6 GHz. Additionally, mmWave channels with the Option 8 split deliver higher SINR while maintaining lower power consumption since the feasibility ratio of this case is higher than the others at high SINR values. After the SINR requirement of 18\,dB, the feasibility of the network drops below 50\%, which is not considered in the figure. Comparing Fig.~\ref{SINR_Na} and Fig.~\ref{PC_REQ}, we observe that in a mmWave fronthaul channel, the Option 8 split achieves high SINR at the UEs while consuming less power at the UAV-APs. 

In Fig~\ref{bar_op}, we observe a significant reduction in power consumption by optimizing the transmit powers and UAV activation for a fair SINR requirement from the max-min fairness algorithm.  Here, we substitute the optimal fair SINR obtained from the max-min fairness algorithm to the SINR requirement $\Gamma_k$ in the power minimization problem. These SINRs are highlighted in the figure. This leads to more energy-efficient UAV-APs maintaining the \ac{QoS} at the UE end.

% In Fig.~\ref{frac_re}, we present the fraction of network realizations that are feasible for achieving optimized transmit power and UAV selection. The SINR requirement at the user is valid only when the network is feasible at least 50\% of the time. 
\textbf{Service time comparison:} The service time for each UAV-AP depends on the battery mass and the total power consumed by the UAV-AP, which includes the mechanical power consumption and communication power consumption. Considering a fixed-wing UAV-AP, the mechanical power consumption includes the weight of the battery and the weight of the aircraft~\cite{7}. Considering the total power consumption and battery energy per unit mass, the service times calculated for Options 8 and 7.2 are 89 minutes and 82 minutes, respectively. We observe the service time of the UAV-AP when
considering split Option 8 at the fronthaul is increased by
8.3\% when compared to split Option 7.2.  

\section{Conclusion}
In this work, we performed joint selection and power optimization in UAV-APs considering the fronthaul capacity limitations in a wireless fronthaul network with a cell-free \ac{mMIMO} access link to UEs. We implemented different functional split options in the fronthaul and analyzed their impact on UE performance and power consumption. Our findings indicate that Option 7.2 outperforms Option 8 in terms of \ac{SINR} at the UEs. In terms of power consumption at the \acp{UAV-AP}, Option 8 proves to be more efficient, and it provides the same SINR as Option 7.2 using more fronthaul bandwidth. 

\bibliography{references}
\bibliographystyle{IEEEtran}

\end{document}